\documentclass[11pt]{article}

\usepackage[margin=1in]{geometry}
\usepackage{amsmath}
\usepackage{newtxtext}
\usepackage{newtxmath}
\usepackage{microtype}
\usepackage{graphicx}
\usepackage{subcaption}
\usepackage{booktabs}
\usepackage[numbers,sort&compress]{natbib}
\usepackage[hidelinks]{hyperref}
\usepackage{authblk}

\providecommand{\Description}[1]{}

\title{\textbf{GenPage: Towards End-to-End Generative Homepage
Construction at Netflix}}

\author[1]{Lequn Wang}
\author[2]{Jiangwei Pan\thanks{Work done while affiliated with Netflix.}}
\author[3]{Fengdi Che \thanks{Work done while the author was an intern at Netflix.}}
\author[4]{Linas Baltrunas}

\affil[1]{Netflix, New York, NY, USA\\
\texttt{lequnw@netflix.com}}
\affil[2]{Netflix, Los Gatos, CA, USA\\
\texttt{panjiangwei@gmail.com}}
\affil[3]{University of Alberta, Edmonton, Alberta, Canada\\
\texttt{fengdi@ualberta.ca}}
\affil[4]{Netflix, Los Gatos, CA, USA\\
\texttt{lbaltrunas@netflix.com}}

\date{}
\begin{document}
\maketitle

\begin{abstract}
We present GenPage, an end-to-end generative approach to Netflix homepage construction that replaces the traditional multi-stage recommender stack with a single transformer. GenPage treats the user and request context as a prompt and autoregressively generates the entire structured, multi-row homepage as the response. We adapt the LLM training recipe: pretraining on production pages, followed by post-training via weighted binary classification (WBC) or reinforcement learning (RL). For industry-scale deployment, we introduce techniques addressing cold start, model freshness, business-rule enforcement, and serving efficiency. In online A/B tests against a mature, highly optimized production homepage recommender, GenPage delivered a substantial lift on the core user engagement metric we use for launch decisions, while reducing end-to-end serving latency by $20\%$. Offline, two findings stand out: enriching the prompt yields a larger improvement than scaling model capacity in our current regime, and RL post-training increases homepage diversity even though diversity is not part of the objective.
\end{abstract}

\maketitle

\section{Introduction}
Large language models (LLMs) have demonstrated a powerful paradigm: a single generative transformer can perform diverse tasks by generating responses conditioned on prompts. More broadly, they have reshaped how we think about machine learning systems: a sufficiently expressive model trained end-to-end can reduce reliance on manually engineered representations and features. 

Traditionally, recommender systems have been built as multi-stage pipelines---candidate generation, ranking, and re-ranking---with each stage optimized separately~\cite{covington2016deep,cheng2016wide}. Constructing a homepage like Netflix's adds another layer of complexity: the output is not a single ranked list but a structured layout of rows and the entities within them. Here, a row is a horizontal collection of entities organized around a theme or purpose (e.g., Continue Watching or Korean TV Shows), and an entity is a movie, show, or other recommendable item. 
The value of a row or entity can depend on the rest of the page. Historically, this layout has been assembled by multiple specialized models, including separate models for rows and entities.

In this paper, we present GenPage, an end-to-end generative approach to Netflix homepage construction. Instead of constructing homepages through multiple stages of specialized models and extensive feature engineering, we train a single generative model to answer a more direct question:
\begin{center}
\begin{minipage}{0.8\linewidth}
\centering
\itshape
Given everything we know about this user and this request, what homepage
should we generate to maximize user satisfaction?
\end{minipage}
\end{center}
It treats the user history and request context as the prompt and autoregressively generates the entire homepage as the response. 

This shift is motivated by several aspirations:
\begin{itemize}
  \item \textbf{End-to-end modeling.} A single transformer model that constructs the page from raw input signals can replace a complex multi-stage recommender stack. This reduces the number of models to maintain, mitigates misaligned objectives across stages, and eliminates much of the traditional feature engineering. 
    \item \textbf{Whole-page optimization via reinforcement learning (RL).} Autoregressive page generation makes it possible to optimize for page-level reward with RL. This can capture interactions across rows and entities, such as diversity or balancing rows with differing \emph{stopping power}---how strongly a row or entity captures attention and stops the user from browsing further.\footnote{For example, a Continue Watching row near the top has high stopping power: users tend to resume a title they have already started rather than explore new content. Visual treatment also matters; rows with larger artwork tend to reduce further scrolling.} Modeling these interactions at the page level lets us align the system more directly with user satisfaction than the entity-level objectives used by most traditional recommender systems.
  \item \textbf{Better scaling behavior.} Recommendation quality can improve with more data, compute, and model capacity without substantial redesign of the system. 
  \item \textbf{Flexibility and extensibility.} The prompt-response paradigm is flexible by design. By simplifying feature engineering and enabling whole-page optimization, GenPage makes it easier to support new product experiences, including additional content types such as live events, games, and podcasts; layouts beyond the current two-dimensional structure; personalized UI components; and per-entity artwork personalization, with fewer architectural changes.
\end{itemize}

Bringing GenPage into production at Netflix also requires addressing several challenges specific to industry-scale recommender systems. Because the homepage is generated in real time, serving latency is a primary engineering constraint. We also need to handle entity cold start in a constantly evolving catalog, keep the model fresh as user interests and trends shift, and enforce complex product and business rules on the generated output. 

Despite these challenges, GenPage has already demonstrated substantial production impact. We validated the weighted binary classification (WBC) post-training variant in an online A/B test against a mature, highly optimized multi-stage production recommender on the Netflix homepage. It delivered a $+0.24\%$ lift on the core user engagement metric we use for launch decisions ($p < 0.001$), one of the largest algorithm-driven gains in recent years, while reducing end-to-end serving latency by $20\%$. Our RL-based post-training has not yet shipped online, but we view it as the key path to realizing GenPage's full potential.

Concretely, this paper makes the following contributions:

\begin{itemize}
  \item We formulate the construction of structured, multi-row personalization surfaces as an end-to-end generative sequence modeling problem, in which a single transformer directly generates the entire page autoregressively. This problem is shared by many industry-scale recommenders, such as Netflix's homepage, Amazon's homepage, and Spotify's home shelves, where the output is a structured layout of rows or modules whose elements interact. This contrasts with prior work on generative recommendation~\cite{rajput2023recommender, geng2022recommendation, tay2022transformer, zhai2024hstu, zhou2025onerec, he2025plum, huang2025rankgpt, agarwal2025pinrec, firooz2025360brew, de2026unified, d2026deploying}, which has primarily focused on generating a flat ranked list of items. Compared with prior work on page-level recommendation~\cite{agarwal2015constrained, wang2016beyond, zhao2018deep, ding2019whole}, GenPage brings the end-to-end generative modeling paradigm of LLMs to whole-page construction, and we address the challenges of doing so at industrial scale in our remaining contributions.
  \item We adapt the LLM training recipe to this setting: pretraining on positively engaged production pages to learn the ``language'' of homepage construction, followed by post-training via weighted binary classification (WBC) or reinforcement learning (RL) to align the pages with user satisfaction. WBC trains the model to estimate the immediate value of each token while still decoding the full homepage autoregressively at serving time; RL optimizes the policy against a page-level reward model trained directly from organic user engagement on each page, in contrast to the pairwise human preferences typically used in LLM RLHF~\cite{christiano2017deep, ouyang2022training}.
  \item We present a set of techniques tailored to industry-scale generative recommendation: a custom tokenization for serving efficiency and product control; context injection and semantic embedding fusion for entity cold start; multi-cadence incremental training for model freshness; constrained decoding for business-rule enforcement; and hybrid row decoding for further latency and cost reduction at inference time.
\item We report offline experiments characterizing the role of pretraining, model scaling, context richness, and RL post-training. Two findings stand out: enriching the prompt yields a larger improvement than scaling model capacity in our current regime, and RL post-training increases homepage diversity even though diversity is not part of the objective.
\item We validate GenPage in an online A/B test against the production homepage recommender at Netflix, demonstrating that it can significantly outperform a mature multi-stage production stack on both user engagement and serving latency.
\end{itemize}
We expect this approach to generalize to many personalization settings. We use Netflix homepage construction as a concrete case study and share our design, trade-offs, and lessons learned.

\begin{figure}[t]
  \centering
  \includegraphics[width=\textwidth]{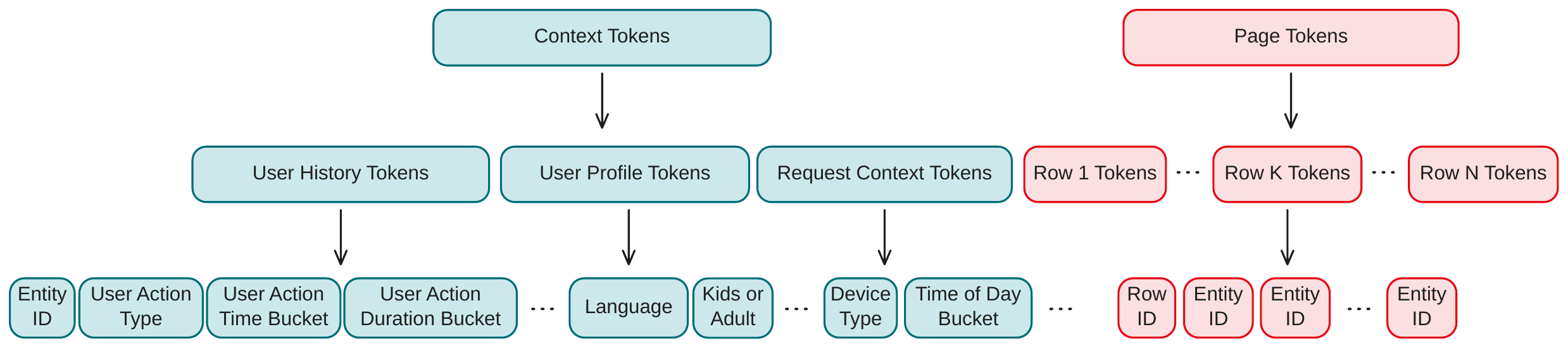}
  \caption{Tokenization of Netflix homepage construction data. The context tokens function as the prompt, drawing from diverse data sources including user history, profile attributes, and request context, with example tokens shown for each source. The page tokens represent the generated response, encoding the structured layout of rows and entities.}
  \Description{Diagram of the tokenization of a Netflix homepage impression. The left portion shows teal context tokens grouped by source---user history, profile attributes, and request context---with example tokens for each source. The right portion shows red page tokens organized as a sequence of rows, where each row begins with a row token followed by its entity tokens, representing the homepage layout that the model generates as output.}
  \label{fig:tokenization}
\end{figure}

\section{Data}
\label{sec:tokenization}
Moving from a traditional recommendation model to a generative transformer requires a fundamental shift in data representation. Just as an LLM represents text as a sequence of tokens, our approach represents both the user context and the generated homepage as a sequence of discrete tokens (Figure~\ref{fig:tokenization}). This sequence includes the full structured homepage layout, with multiple rows and the entities inside them, so the model can generate the page holistically rather than scoring each row or entity in isolation.

Each training example represents a homepage impression and consists of three components:
\begin{itemize}
    \item \textbf{Context}: user engagement history, profile attributes, and request context.
    \item \textbf{Page}: the recommended rows and entities shown on the homepage, in layout order.
    \item \textbf{Feedback}: user interactions with that page, such as play, thumbs-up, and abandonment for entities on the page.
\end{itemize}

Only the context and page are tokenized as model inputs and outputs. Feedback is used to derive supervision signals via our internal reward system (Section~\ref{sec:reward}).

Instead of using an off-the-shelf text tokenizer, we build a domain-specific tokenizer for the homepage construction data. This is a proven approach in recommender systems~\cite{kang2018self} and other specialized domains including computer vision~\cite{ramesh2021zero}, biology~\cite{rives2021biological}, and chemistry~\cite{schwaller2019molecular}. Compared with generic text tokenization, this gives us two key advantages:

\begin{itemize}
  \item \textbf{Computational efficiency.} Custom tokenization significantly reduces sequence length, thus lowering inference cost and latency. For example, representing the event ``User watched Orange Is the New Black for 50 minutes 30 days ago.'' would require 16 tokens with GPT-5, whereas our scheme compresses it to 4 tokens: \texttt{[Entity\_ID]}, \texttt{[Action\_Type]}, \texttt{[Action\_Time\_Bucket]}, and \texttt{[Action\_Duration\_Bucket]}.
  \item \textbf{Product control.} A direct mapping between tokens and product concepts, such as rows and entities, makes it easier to control what the model can generate. This is crucial for enforcing business rules on the final homepage (Section~\ref{sec:business_rules}).
\end{itemize}

\subsection{Context Tokens}
Context tokens encode user engagement history, user profile, and request context. 

We represent user history as a sequence of tokenized user actions~\cite{hidasi2016session, kang2018self, sun2019bert4rec}. For each action, we extract key metadata as tokens, including action type, entity ID, timestamp, and duration. These actions include both explicit signals, such as play, add to My List, and thumbs-up, and implicit signals, such as trailer views or visits to a details page.

User profile tokens capture attributes such as language and profile type. Request context tokens encode signals like time of day, day of week, and device.

Some data sources are too long to include directly as raw token sequences. A user's full impression history, for example, would be prohibitively expensive to tokenize in full. In such cases, we use a summarized version. This is a pragmatic trade-off: while GenPage aims to operate on raw inputs as much as possible, handcrafted summaries still introduce a form of prompt engineering into the pipeline. Learning to compress these long data sources end to end is an important direction for future work. 

To help the model distinguish between data sources, we insert special tokens that mark the start of each source segment. Continuous signals, such as timestamps and durations, are bucketized into discrete ranges to maintain a finite vocabulary.

\subsection{Page Tokens}

We represent each entity (e.g., a movie or show) and each row (e.g., Korean TV Shows) as a single token. The homepage is tokenized in layout order: left to right, top to bottom. We update the entity and row vocabulary daily to incorporate newly added entities and rows. At serving time, entities that are still out of vocabulary are handled through semantic embedding fusion (Section~\ref{sec:cold_start}) and fallback tokens (Section~\ref{sec:incremental}). 

In principle, the same paradigm can extend to any output that can be expressed as a linear token sequence. This includes layouts beyond the current two-dimensional structure, such as one-dimensional feeds or mixed layouts, as well as personalized UI components, such as the display size of each row, and per-entity outputs such as personalized artwork. We leave these extensions to future work.

\subsection{Paginated Recommendation}
\label{sec:pagination}

To make recommendations responsive to in-session user preferences, the homepage is often generated incrementally, a few rows at a time. Before each pagination request, we append the page tokens from previously generated rows to the prompt, along with the latest user engagements on those rows, derived from Netflix's real-time event-logging infrastructure. This allows the model to generate the next set of recommendations using both the user's long-term preferences and their most recent in-session engagements.

\section{Reward System}
\label{sec:reward}

To quantify the long-term value of a recommendation, we rely on an internal reward system described in prior work~\cite{tang2023reward}. This reward system is tuned through online A/B testing to align with long-term user satisfaction and serves as the core supervision signal for both supervised and reinforcement learning.

The reward system processes user feedback and assigns a scalar reward for every impressed entity on the homepage. For instance, a TV show binge-watched in one night reflects stronger user satisfaction and receives a higher reward than a movie watched for only 10 minutes. An impressed but abandoned entity receives a negative reward. 

We define the page-level reward as the sum of rewards across all impressed entities on the homepage.

\section{Model Architecture}
\label{sec:architecture}

GenPage uses a standard decoder-only transformer~\cite{radford2019language}, the same general architecture that underlies many modern LLMs. This choice keeps the model simple and flexible, while allowing us to leverage advances in transformer architectures, training methods, and serving systems from the broader LLM ecosystem.

One architectural detail is that we untie the input embedding and output projection weights~\cite{inan2016tying,press2017using}. This is useful because pretraining and post-training place different demands on the logits. The optimal logit scales differ substantially between next-token-prediction pretraining (which optimizes a softmax over the vocabulary) and WBC post-training (which optimizes per-token sigmoids); see Section~\ref{sec:training} for both. Untying the weights gives the model more flexibility to adapt to both objectives.

For our first rounds of online A/B tests, we use a $\sim 200$M-parameter model to ensure we stay within our serving latency budget. The offline scaling trends (Section~\ref{sec:offline_evaluation}) suggest further quality improvements are available as inference optimizations create latency headroom.

\section{Training Recipe}
\label{sec:training}
Our training pipeline mirrors the LLM recipe: we first teach the model the ``language'' of the Netflix homepage through pretraining, then align its outputs with user satisfaction through post-training. For post-training, we explore two alternative approaches: weighted binary classification (WBC) and reinforcement learning (RL). 

WBC is simpler to optimize and aligns directly with the entity-level objectives of our production ranking models. RL is harder to evaluate and optimize, but it is the key path to GenPage's full vision of page-level optimization, with the flexibility to incorporate test-time reasoning and multi-token entity representations.

\subsection{Pretraining via Next-Token Prediction}

We pretrain the model with a standard next-token-prediction objective: given the context tokens and a prefix of page tokens, the model learns to predict the next page token. This stage focuses on representation learning, teaching the model the relationship between user contexts and successful homepages. Note that our context-page training examples resemble the prompt-response pairs used in LLM supervised fine-tuning (SFT) more than the raw text used in LLM pretraining. We nonetheless call this stage \emph{pretraining} because we train the model from scratch rather than fine-tuning from an existing checkpoint. 

Unlike LLMs, which often face a scarcity of high-quality labeled data, recommender systems have an abundance of user feedback. For pretraining, we use homepage impressions that received positive user feedback when served in production, bootstrapping the model to generate pages similar to those produced by the production system.

However, pretraining mainly teaches GenPage to imitate the production system. It does not directly optimize the magnitude of the reward, and as GenPage becomes part of production, repeatedly training on pages generated by earlier versions of the model risks model degeneration~\cite{shumailov2024collapse}. To address these limitations, we explore two post-training approaches.

\subsection{Post-Training via Weighted Binary Classification}
\label{sec:wbc}
One effective way to align the generative model with user satisfaction is weighted binary classification (WBC). At a high level, WBC turns generation into token-level value prediction: given the user context and the tokens generated so far, the model learns to estimate the value of generating each possible next row or entity token. 

This objective is easier to optimize than page-level RL. By decomposing the homepage into per-token targets, WBC provides token-level credit assignment by construction, rather than requiring RL to infer how each generated decision contributed to the final page-level reward.

This training setup is enabled by our custom tokenization (Section~\ref{sec:tokenization}). Each page token corresponds directly to a specific entity or row, making it straightforward to assign a reward. For every impressed entity on the page, our reward system provides a scalar reward based on user feedback. For each impressed row, we derive a row-level reward by aggregating the rewards of the entities in that row.

From each reward, we derive a binary label from its sign, such as positive engagement versus abandonment, and a weight from its magnitude, such as binge-watching receiving a higher weight than a short play. We then optimize a weighted binary cross-entropy loss on the logit for the corresponding token. Under this setup, the logit for a token can be interpreted as the model's value estimate for generating that token at that position.

For each impressed entity and row target, we additionally sample random tokens as negative targets. Because rarely-impressed tokens are drawn as random negatives far more often than they appear as positives, this injects a mild pessimism~\cite{swaminathan2015batch,kumar2020conservative}---effectively a popularity prior---into their learned values, and prevents the model from confidently surfacing niche entities whose high scores would otherwise be dominated by noise.

Despite being trained as a value predictor, the model can still generate pages autoregressively. At each step, it computes the values (logits) for all candidate tokens, greedily selects the one with the highest value, and appends it to the prefix. This process repeats token by token to generate the entire homepage. 

On our standard held-out evaluation set for production ranking models, the WBC-trained generative model outperforms our production rankers on key entity-level metrics such as weighted AUC. This demonstrates that an end-to-end generative formulation can match or exceed a multi-stage production stack on standard entity-level offline metrics.

\begin{figure}[t]
  \centering
  \includegraphics[width=\textwidth]{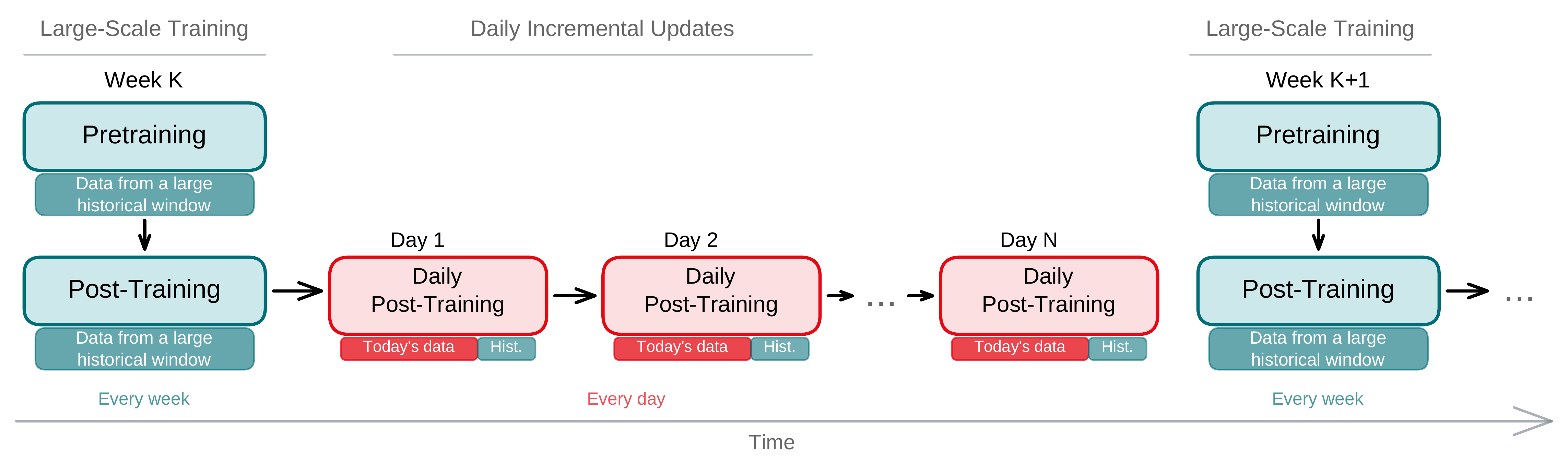}
  \caption{Multi-cadence incremental training strategy. Periodic large-scale pretraining and post-training passes run on a broad historical window. Between them, daily incremental post-training updates combine the latest day's data with a sampled subset of past data to keep the model fresh while avoiding catastrophic forgetting.}
\Description{Diagram of a cyclic training schedule with two cadences. Left and right sections show large-scale training cycles (Week K, Week K+1) with sequential Pretraining and Post-Training blocks, each drawing from a broad historical window (teal bars). The middle section shows daily Post-Training updates over Day 1, Day 2, ..., Day N: each combines today's data (red bar, the dominant share) with a sampled subset of historical data (teal bar). A time arrow runs along the bottom; color-coded labels mark the two cadences (``Every week'' in teal, ``Every day'' in red).}
  \label{fig:incremental}
\end{figure}

\subsection{Post-Training via Reinforcement Learning}
\label{sec:rl}
Reinforcement learning (RL) is the post-training direction we are actively exploring. While WBC is effective at optimizing entity-level metrics, it does not directly optimize the homepage as a whole. By treating page generation as a sequential decision-making process, RL enables whole-page optimization while preserving the flexibility of autoregressive generation. This can open the door to several important capabilities: 

\begin{itemize}
  \item \textbf{Whole-page optimization.} RL directly optimizes an aggregate page-level reward, accounting for interactions across rows and entities, such as diversity and stopping power, as well as broader page-level business constraints.
  \item \textbf{Test-time reasoning.} Analogous to its application in LLMs, RL can optimize reasoning capabilities for generative recommendation~\cite{tsai2024leveraging, liang2026generative}. Reasoning outputs can also be viewed as a form of automated feature engineering.
  \item \textbf{Multi-token entity support.} In our custom tokenization, each entity and row is a single token, so rewards map cleanly to individual tokens. In more complex settings, however, an entity may be represented by multiple tokens (e.g., \texttt{[Show\_ID]} + \texttt{[Episode\_\#]} for an episode within a show, or multi-token semantic IDs~\cite{rajput2023recommender}). In that case, WBC's per-token labeling becomes ambiguous because a single entity-level reward must be distributed across multiple tokens. RL avoids this issue by optimizing the sequence-level return, making it a more natural fit for variable-length, multi-token entities.
\end{itemize}

Of these, only whole-page optimization is explored in this paper; test-time reasoning and multi-token entity support are part of the motivation for adopting RL but are directions for future work.

Inspired by the two-step RLHF recipe~\cite{christiano2017deep, ouyang2022training} used to align large language models, we first train a reward model that predicts the page-level outcome reward for a generated page and then optimize the generation policy against its predictions. This reward model is distinct from the reward system of Section~\ref{sec:reward}. The reward system converts \emph{observed} user feedback into a scalar reward for a page that was actually shown, whereas the reward model \emph{predicts} the page-level reward for any generated page without showing it to a user. This prediction is what lets RL optimize against arbitrary candidate pages during training. 

Training against a reward model avoids the high variance of off-policy correction on logged or predicted propensities~\cite{chen2019top}, but introduces the risk of reward hacking. Since the reward model is trained on data generated from the production policy, it is most reliable on pages similar to those the production policy generates. We therefore use a KL penalty to keep the policy close to the pretrained checkpoint, which itself was trained to mimic the production policy. This keeps the pages within the reward model's region of coverage and limits opportunities for reward hacking.

For the RL algorithm, we adopt Dr.\ GRPO~\cite{liu2025understanding}, a variant of GRPO~\cite{guo2025deepseek} that mitigates biases in the training objective. We build our training pipeline using the verl RL library~\cite{sheng2025hybridflow}, with vLLM~\cite{kwon2023efficient} as the inference engine. To train the model within this framework, we need the following components:

\begin{itemize}
    \item \textbf{Prompts}: Production user requests, represented by context tokens.
    \item \textbf{Policy and reference models}: Both are initialized from the pretrained checkpoint; the reference model anchors the KL penalty discussed above.
    \item \textbf{Reward model}: A dedicated transformer-based reward model, also initialized from the pretrained checkpoint, predicts the page-level outcome reward, using the sum of entity-level rewards from our internal reward system as the supervision target. We also incorporate rule-based rewards to guide the RL policy. For example, the page should resemble a list of rows, and business-critical rows or entities should not appear too low on the page.
\end{itemize}

\subsubsection*{Open Challenges}

Unlike LLM RLHF, which relies on scarce, human-labeled pairwise preference comparisons over unstructured text responses, our setting has direct user feedback at the entity and row level on a structured page. Fully leveraging this richer, structured signal raises two related challenges. (1) \emph{Offline evaluation}: our current page-level evaluation relies on a reward model the policy is also optimizing against, making it largely circular. Human labels---a common alternative in LLM evaluation---are also problematic, since each user's preferences depend on personal context that third-party labelers cannot replicate. Candidate directions include dedicated held-out reward models, rule-based evaluations, and counterfactual estimators~\cite{bottou2013counterfactual,swaminathan2017off, joachims2021recommendations}. (2) \emph{Decomposition vs.\ joint modeling}: WBC and page-level RL sit at opposite extremes: WBC fully decomposes the page into entity-level optimization (simplifying optimization but losing some interactions across rows and entities), while page-level RL optimizes the entire page jointly without structural decomposition (capturing all interactions across rows and entities but being harder to optimize and less sample-efficient). Middle grounds that exploit partial structure---mixed objectives, structured policy parameterizations, or assumptions from user behavior models (e.g., click models~\cite{chuklin2022click} or choice models~\cite{train2009discrete}), especially those developed for 2D page layouts~\cite{kang2025rethinking, kang2026carousel}---are promising.

\section{Addressing Production Challenges}
\label{sec:challenges}
\subsection{Cold Start}
\label{sec:cold_start}
New entities lack the rich interaction data needed to learn robust token embeddings. We address this through two complementary strategies:

\begin{itemize}
  \item \textbf{Context injection.} We inject metadata about new or time-sensitive entities (e.g., Live Now events) directly into the context tokens, providing the model with semantic and time-sensitive information.
  \item \textbf{Semantic embedding fusion.} Rather than relying solely on entity ID embeddings learned from user interaction data, we represent each entity as a fusion of its ID embedding and a content-based embedding derived from semantic information such as synopses, cast, transcripts, genres, and video content. This fused embedding serves as the input embedding for the entity's token in the transformer. During training, with a small probability, we randomly replace an entity ID token with the generic fallback token (Section~\ref{sec:incremental}), so the model learns to make recommendations from the content-based embedding alone. This ensures that a new entity has a meaningful representation in the same latent space as established entities as soon as its content metadata is available, even before it has any interaction data. 
\end{itemize}

\subsection{Multi-Cadence Incremental Training}
\label{sec:incremental}

At Netflix scale, daily retraining of a large transformer from scratch is prohibitively expensive, but recommendation models must remain fresh to capture shifting trends and new catalog additions. We address this with a multi-cadence incremental training strategy (Figure~\ref{fig:incremental}).

Our training pipeline operates on a cyclic schedule with two distinct rhythms. At a tunable cadence, we conduct a large-scale pretraining and post-training pass on data from a broad historical window. Between these passes, each day we perform an incremental update by continuing post-training from the previous day's checkpoint, using a mix of the latest day's data and a sampled subset of past data. This helps the model stay current with new trends and catalog changes while preventing overfitting and catastrophic forgetting~\cite{kirkpatrick2017overcoming}.

To manage the daily influx of new tokens (e.g., new entities, rows), we employ fallback tokens. New tokens are initialized using fallback tokens of their type (e.g., \texttt{[Row\_Fallback\_Token]} for new rows, \texttt{[Entity\_Fallback\_Token]} for new entities). During training, we randomly replace a small percentage of known tokens with fallback tokens, teaching the model to handle unknown tokens gracefully.

\subsection{Enforcing Business Rules}
\label{sec:business_rules}
A Netflix homepage must satisfy structural constraints (e.g., organized as a list of rows) as well as product logic such as deduplication, row pinning, and category consistency (e.g., entities in a Comedy row must be comedies).  While training signals can encourage rule adherence, they cannot guarantee strict compliance. 

We enforce these rules at inference time through \emph{constrained decoding}. At each autoregressive generation step, we compute a mask of eligible tokens based on the applicable business rules and apply it to the output logits, allowing only rule-compliant tokens to be generated. This is greatly simplified by our custom tokenization (Section~\ref{sec:tokenization}): because each entity and row is a single token, business rules map directly to token-level masks, avoiding the multi-token bookkeeping that constrained decoding requires over a text vocabulary. For example, to pin a specific row (e.g., popular games) at a fixed position (e.g., row position 2), we simply mask out all other tokens at that position.  

A challenge for GenPage is that the applicable business rules can vary across requests. Unlike common LLM constrained-decoding settings that reuse a fixed set of constraints, GenPage must construct token masks dynamically for each request and each position, requiring several customizations to the inference engine to keep decoding latency within budget.

\subsection{Hybrid Row Decoding}
\label{sec:hybrid_decoding}

Autoregressive generation ensures that each newly generated token is conditioned on the full preceding context, but generating every entity token one at a time can be expensive. We leverage the structure of the homepage to balance inference efficiency with the amount of contextual information available to each generated token. 

Within each row, the first few entities are especially important: they receive the most user attention and strongly shape the row's perceived quality and theme. To reduce inference latency, we use a hybrid row decoding strategy. The model autoregressively generates only the first few entities in each row. Conditioned on this generated prefix, we obtain logits for all eligible entities in a single forward pass and select the top-scoring remaining entities, subject to the same inference-time business-rule constraints described above. 

This approach preserves autoregressive conditioning where it matters most while avoiding the latency and cost of decoding long rows token by token.

\section{Offline Experiments}
\label{sec:offline_evaluation}

We present ablations on Netflix internal data characterizing how different components of GenPage affect model quality. Because the system was developed iteratively, individual ablations span different training configurations and data snapshots, so we report only relative comparisons within each study. Unless otherwise noted, experiments use $\sim 200$M-parameter models and report results on a held-out evaluation set. We focus on metrics and findings most relevant to practitioners.

\begin{table}[t]
\centering
\caption{WBC post-training performance with and without pretraining. Loss is the weighted binary cross-entropy. Row AUC and Entity AUC are sample-weighted ROC-AUC over row/entity targets.}
\label{tab:pretraining-results}
\begin{tabular}{lcc}
\hline
\textbf{Metric} & \textbf{With Pretraining} & \textbf{Without Pretraining} \\
\hline
Loss       & \textbf{0.321} & 0.333 \\
Row AUC    & \textbf{0.884} & 0.879 \\
Entity AUC & \textbf{0.920} & 0.910 \\
\hline
\end{tabular}
\end{table}

\subsection{Does pretraining help?}

We compare WBC post-training with and without a preceding next-token-prediction pretraining stage. Table~\ref{tab:pretraining-results} shows that pretraining yields substantial improvements across all metrics. The gains may look small in absolute terms, but they are large in our production regime: setting aside the sample weighting, an Entity AUC lift from $0.91$ to $0.92$ means that for a randomly drawn positive-negative pair of impressed entities, the model's misranking rate drops from $9\%$ to $8\%$---a magnitude of improvement we rarely observe from a single change on a mature production system. Pretraining the model on the ``language'' of the Netflix homepage provides a strong initialization for post-training, mirroring the pretrain-then-post-train recipe behind LLMs.

\subsection{How does the performance scale with model size?}

\begin{figure}[t]
  \centering
  \includegraphics[width=0.65\linewidth]{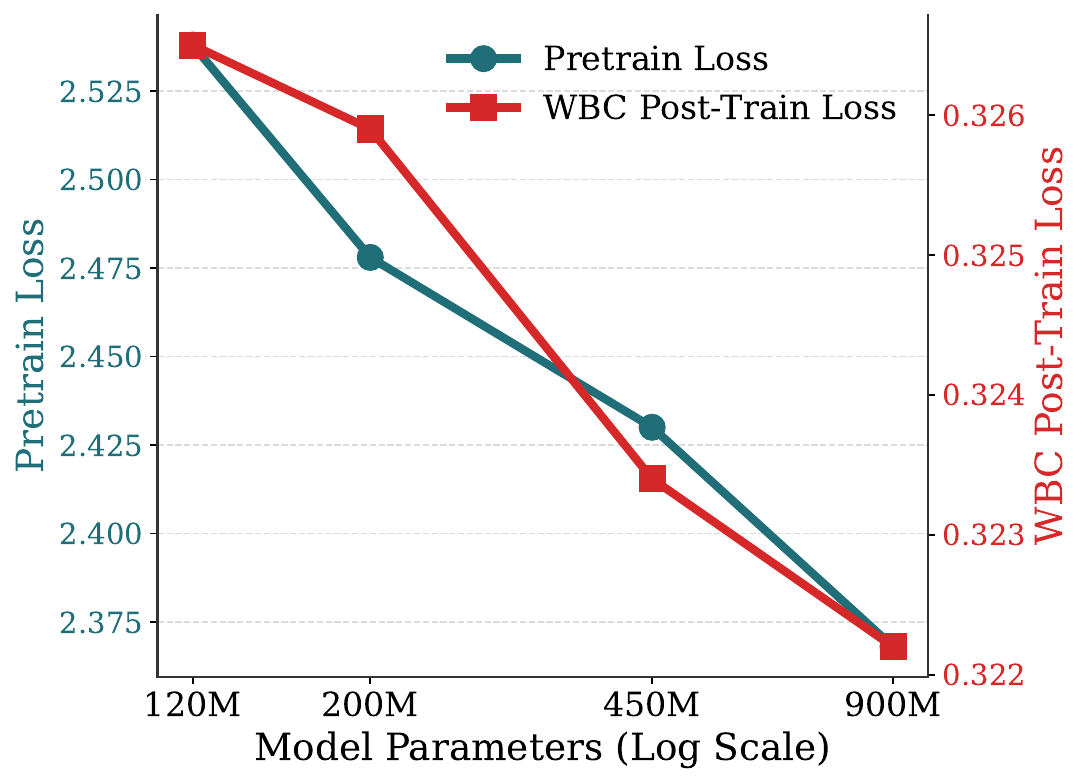}
  \caption{Pretraining and WBC post-training losses as model size scales from 120M to 900M parameters. Both decrease in a power-law-like fashion, mirroring LLM scaling trends.}
  \Description{Line chart with model size on the x-axis (from approximately 120M to 900M parameters, log scale) and loss on the y-axis. Two lines are plotted: the pretraining next-token-prediction loss and the WBC post-training loss. Both decrease in a power-law-like fashion as model size increases.}
  \label{fig:exp_model_size}
\end{figure}

We sweep model size from $\sim 120$M to $\sim 900$M parameters (Figure~\ref{fig:exp_model_size}) and report the next-token-prediction loss from pretraining and the WBC loss from post-training. Both losses decrease in a power-law-like fashion, mirroring the scaling trends seen in LLMs~\cite{kaplan2020scaling, hoffmann2022training}. This confirms that the generative approach scales favorably with model size, suggesting that recommendation quality can be improved by scaling capacity.

\subsection{How does the performance scale with information in user context?}

\begin{figure}[t]
  \centering
  \includegraphics[width=0.65\linewidth]{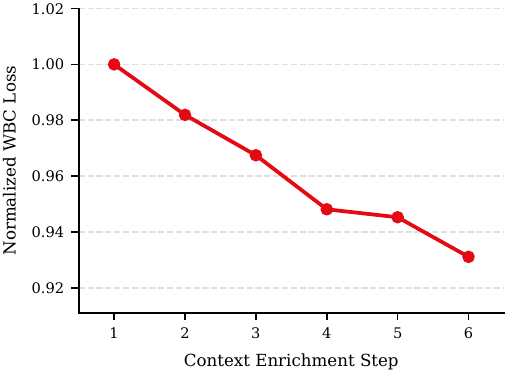}
  \caption{WBC post-training loss as we progressively enrich the user context tokens. Loss is normalized to step 1 (= 1.0). Since individual experiments were conducted on different data snapshots, each step's value is compounded from the measured relative improvement over its predecessor rather than measured directly against a shared baseline.}
  \Description{Line chart showing WBC post-training loss across six progressive enrichment steps, with model size held fixed. The y-axis is loss normalized to step 1 (= 1.0); loss decreases monotonically to approximately 0.93 at step 6. The cumulative reduction is substantially larger than the improvement from scaling the model from 120M to 900M parameters.}
  \label{fig:exp_enrich_context}
\end{figure}

Over the course of development, we progressively enriched the prompt, both by adding new data sources to the context and by refining how each source is tokenized. With model size held fixed, Figure~\ref{fig:exp_enrich_context} shows that the WBC post-training loss decreases substantially as the context is enriched.

The model-size sweep and the context-enrichment sweep span different axes and are not strictly comparable: the model-size study covers roughly an order of magnitude in parameters, while the context study spans the full trajectory of our prompt design. Even so, the gap between the two is striking. Scaling the model from $120$M to $900$M parameters reduces WBC loss by roughly $1.3\%$, whereas the cumulative effect of enriching the context is around $6.9\%$. In several cases, a single well-designed context addition delivers a larger improvement than the entire $\sim 7.5\times$ model capacity scaling.

This suggests that, in our regime, enriching the prompt---both what we put in the context and how we tokenize it---yields a substantially larger improvement than scaling model capacity. Personalization quality appears to be bottlenecked first by the information and representation available to the model, and only then by capacity. We expect context enrichment to dominate until the context is saturated, at which point model capacity becomes the primary driver. 

\subsection{Does RL post-training optimize at the page level?}

\begin{figure}[t]
  \centering
  \includegraphics[width=0.65\linewidth]{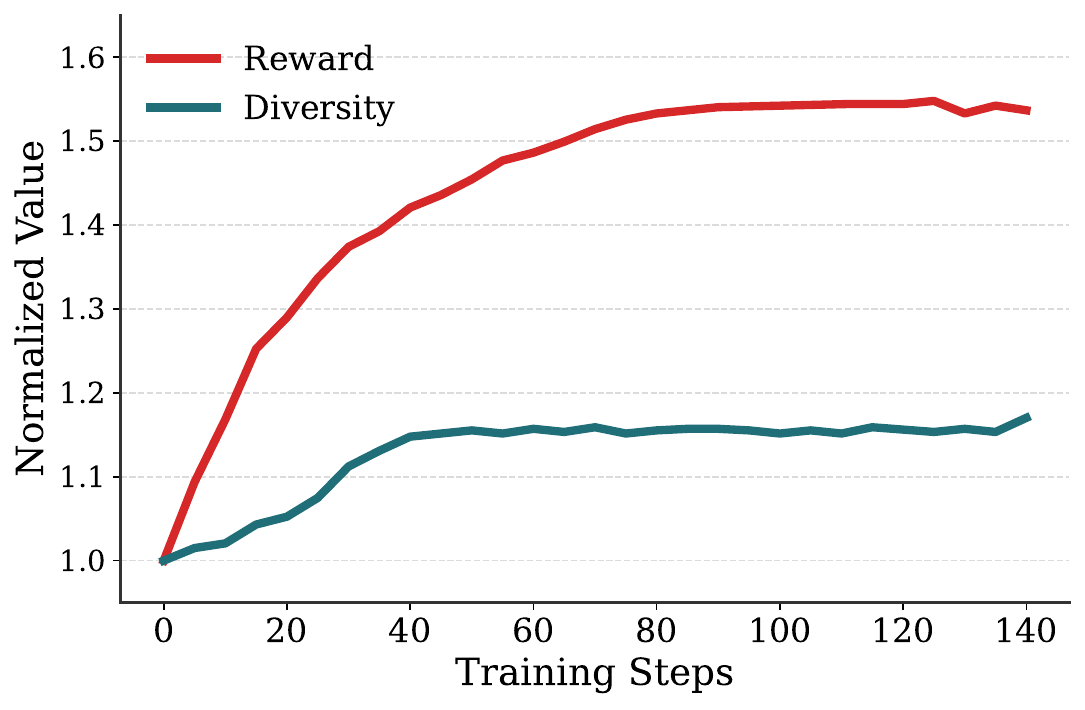}
  \caption{Training dynamics for RL post-training. Reward and diversity are shown relative to the initial pretrained checkpoint (1.0). Reward steadily increases as expected. Diversity also increases substantially despite not being part of the RL objective, suggesting that the policy is optimizing the page as a whole rather than myopically optimizing each token in isolation.}
  \Description{Line chart of two metrics over 140 RL training steps, normalized so the initial pretrained checkpoint is 1.0. Reward (red line) rises from 1.0 to approximately 1.55, plateauing around step 80. Diversity (teal line) rises from 1.0 to approximately 1.17, plateauing around step 50, despite not being part of the RL objective.}
  \label{fig:rl_training}
\end{figure}

In offline evaluations (Figure~\ref{fig:rl_training}), RL post-training consistently improves the page-level reward over the pretrained checkpoint, but this is largely confirmatory: the reward is computed using the same model the policy is optimizing against. More interestingly, although diversity is not part of the RL objective, homepage diversity---measured via pairwise embedding distance among entities on the page---also increases over the course of training. This suggests that the RL-trained policy is optimizing the page as a whole rather than myopically optimizing each token in isolation. 

\section{Online Evaluation}
\label{sec:online_evaluation} 

\begin{figure}[t]
  \centering
  \includegraphics[width=0.65\linewidth]{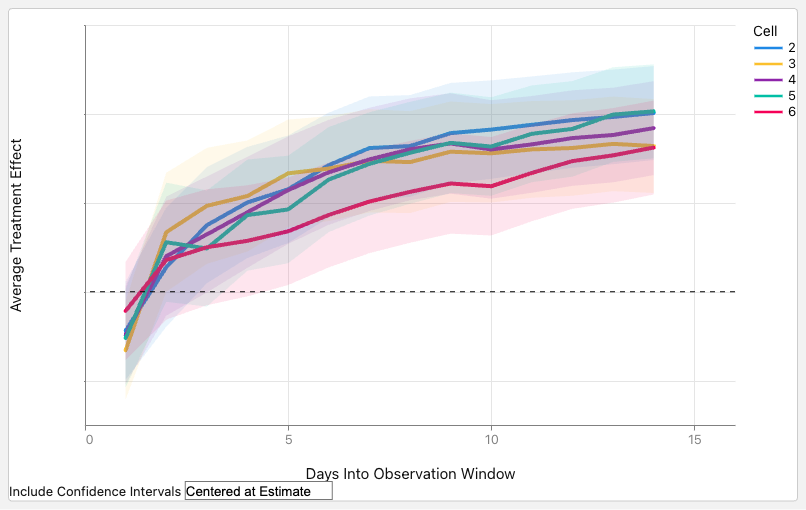}
  \caption{Daily core user engagement metric over a 14-day online A/B test. The figure plots the average treatment effect of five GenPage variants (cells 2--6, differing in training-data configurations) against the production baseline (cell 1) on the metric we use for launch decisions. Shaded regions indicate $95\%$ confidence intervals. All five variants delivered statistically significant improvements over production, with the best-performing variant (cell 2) reaching a $0.24\%$ relative lift by day 14. }
  \Description{Line chart of the daily average treatment effect over a 14-day period for five GenPage variants (cells 2 through 6) against the production baseline (cell 1) on a core user engagement metric. Each variant is plotted as a separate line with a shaded $95\%$ confidence interval. All five variants remain above zero throughout the 14-day period, indicating statistically significant positive lift over the baseline.}
  \label{fig:online_evaluation}
\end{figure}

We conducted an online A/B test against the current production homepage recommender using a $\sim 200$M-parameter WBC model. In this test, GenPage decoded over the existing production row and entity candidate sets, which help handle many business rules (e.g., eligibility). We evaluate only the WBC variant online; scaling RL training to our data volume, refining the page-level reward model, and strengthening its offline evaluation are ongoing, and we plan to evaluate RL post-training in subsequent A/B tests. 

Figure~\ref{fig:online_evaluation} shows the result: all five variants delivered statistically significant improvements on the core user engagement metric we use for launch decisions ($p < 0.001$) against a mature, highly optimized multi-stage production baseline. The best variant delivered a $+0.24\%$ lift (95\% CI $[0.17\%, 0.30\%]$). This represents one of the largest algorithm-driven gains in recent years. The variants explored different amounts of random negatives (Section~\ref{sec:wbc}) and different filters on which production impressions to include in training; that all five delivered comparable lifts suggests the gain is robust to these design choices rather than dependent on a particular configuration.

Alongside the engagement wins, we observed unintended shifts in the mix of impressed entities across several dimensions (e.g., new vs.\ established titles, TV shows vs.\ movies, and higher vs.\ lower engagement quality).
These shifts are not necessarily negative, but they are not something we explicitly optimized for, and they warrant deeper investigation. We suspect these shifts reflect GenPage personalizing more precisely than the production stack---consistent with an increase in homepage impression efficiency, i.e., users engaging with what they saw using fewer impressions. This sharper personalization appears to surface components inherited from the production system (such as the reward system) that aren't yet aligned with the new generative paradigm. We plan to characterize the drivers of these shifts and, where appropriate, tune these components so the resulting distributions better align with desired product behavior.

We also observed strong responsiveness to in-session signals: the latest in-session actions (Section~\ref{sec:pagination}) quickly influenced subsequent recommendations and faded back to long-term preferences after a day or two, confirming that the model effectively attends to action timestamps. This responsiveness emerges naturally from the generative formulation, without the extensive manual feature engineering traditionally used in our production stack.

Contrary to the common assumption that generative models are slower, GenPage reduced end-to-end serving latency by $20\%$ relative to the baseline. By replacing multiple ranking stages and heavy feature computation with a single transformer model operating on raw tokenized inputs, we eliminated substantial serving complexity and computational overhead. Custom tokenization and hybrid row decoding further reduced the number of decoding steps and thus latency. The $20\%$ reduction was achieved without exhausting the available optimizations; further reductions are possible. This headroom can be reinvested in capacity or richer prompts.

\section{Conclusion}
We presented GenPage, an early step toward end-to-end generative Netflix homepage construction: representing user context as a tokenized prompt and generating the entire homepage autoregressively in real time. This collapses the multi-stage recommender stack into a single transformer that can be optimized end-to-end. 

In online A/B tests against a mature, highly optimized multi-stage production system, GenPage delivered a substantial lift of $+0.24\%$ on the core user engagement metric we use for launch decisions ($p < 0.001$), while reducing end-to-end serving latency by $20\%$. Achieving this required adapting the LLM training recipe---pretraining followed by WBC or RL post-training---together with a set of domain-specific techniques: custom tokenization for serving efficiency and product control, context injection and semantic embedding fusion for entity cold start, multi-cadence incremental training for model freshness, constrained decoding for business-rule enforcement, and hybrid row decoding for inference efficiency. 

Two offline findings stand out. First, in our current regime, enriching the prompt yields a substantially larger improvement than scaling model capacity---a takeaway we expect to generalize to other industry-scale personalization settings, at least until the available context is fully exploited. Second, RL post-training increases homepage diversity even though diversity is not part of the objective---an indication that page-level optimization captures interactions across rows and entities. 

Several pieces of the full vision are still in progress: RL post-training has not yet shipped online, long context still relies on handcrafted summarization, and broader LLM-style capabilities---language, multimodality, and reasoning---have not yet been incorporated. One promising direction here is a hybrid tokenization combining our domain-specific tokens with generic text tokens, retaining structured control while inheriting the strengths of general-purpose LLMs; conceptually, this introduces an additional recommendation modality into an LLM.

More broadly, we expect many advances from the LLM ecosystem to transfer naturally to this setting, and the boundary between an LLM and a recommender system may increasingly blur. Our results suggest this is a viable path toward simpler, more flexible recommender systems that align more directly with user satisfaction and can more readily support new product experiences.

\section*{Acknowledgments}
We thank, in alphabetical order, Abhishek Agrawal, Ashish Rastogi, Baolin Li, Casey Stella, Dan Zheng, Daneo Zhang, Ding Tong, Donnie DeBoer, Fernando Amat Gil, Grace Huang, Hakan Baba, Inbar Naor, Ishita Verma, Jason Uh, Jimmy Patel, Justin Basilico, Lanxi Huang, Lingyi Liu, Liping Peng, Louis Wang, Michelle Kislak, Nathan Kallus, Nicolas Hortiguera, Paran Jain, Qusai Al-Rabadi, Rein Houthooft, Ryan Lee, Santino Ramos, Scarlet Chen, Shaojing Li, Sheallika Singh, Si Cheng, Wei Wang, Yesu Feng, and ZQ Zhang for their contributions to this work. 

\bibliographystyle{unsrtnat}
\bibliography{references}










\end{document}